\RequirePackage{fix-cm}
\documentclass[twocolumn]{svjour3}          % twocolumn
\smartqed  % flush right qed marks, e.g. at end of proof
\usepackage{graphicx}
\usepackage{makeidx}
\usepackage[utf8]{inputenc}
\usepackage{amsmath}
\usepackage{amsfonts}
\usepackage{amssymb}
\usepackage [english]{babel} 
\usepackage{mathrsfs}
\usepackage{graphicx}
\usepackage{subfigure}
\usepackage{theorem}
\begin{document}

\title{Gravitational entropy of black holes and wormholes
}


\author{Gustavo E. Romero \and Romain Thomas \and Daniela P\'erez}


\institute{Gustavo E. Romero \and Daniela P\'erez \at
              Instituto Argentino de Radioastronom{\'\i}a, C.C.5, (1984)\\
              Villa Elisa, Bs. As., Argentina \\
              \email{romero@iar-conicet.gov.ar,\\
               danielaperez@iar.unlp.edu.ar}           %  \\
           \and
           Romain Thomas \at
           Universit\'e Paris-Sud 11, 91400 Orsay, France\\
           \email{romain.thomas@u-psud.fr}}

%\author{First Author         \and
%        Second Author %etc.
%}


%\institute{F. Author \at
%              first address \\
%              Tel.: +123-45-678910\\
%              Fax: +123-45-678910\\
%              \email{fauthor@example.com}           %  \\

\date{Received: date / Accepted: date}

\maketitle

\begin{abstract}

Pure thermodynamical considerations to describe the entropic evolution of the universe seem to violate the Second Law of Thermodynamics. This suggests that the gravitational field itself has entropy. In this paper we expand recent work done by Rudjord, Gr$\varnothing$n and Sigbj$\varnothing$rn where they suggested a method to calculate the gravitational entropy in black holes based on the so-called `Weyl curvature conjecture'. We study the formulation of an estimator for the gravitational entropy of Reissner-Nordström, Kerr, Kerr-Newman black holes, and a simple case of wormhole. We calculate in each case the entropy for both horizons and the interior entropy density. Then, we analyse whether the functions obtained have the expected behaviour for an appropriate description of the gravitational entropy density.
\keywords{Gravitation \and General Relativity \and entropy \and Weyl tensor \and black holes \and wormholes}

\end{abstract}

%\section{Introduction}
\section{Introduction}
\label{intro}
The world changes. This basic fact is expressed in the Second Law of Thermodynamics: {\em The entropy of a closed system never decreases}:
\begin{equation}
	\frac{dS}{dt}\geq 0,
\end{equation}
where the entropy is denoted by $S$.

In the 1870s, Ludwig Boltzmann \cite{Boltz1}\cite{Boltz2} provided a description of the entropy relating the states of macro and micro-systems. He argued that any system would evolve toward the macrostate that is consistent with the largest possible number of microstates. The number of microstates and the entropy of the system are related by the fundamental formula:
%\paragraph{}
\begin{equation}
S= k_{\rm{B}} \ln W,
\end{equation}
where $k_{\rm{B}}=10^{-23}$ JK$^{-1}$ is Boltzmann's constant and $W$ is the volume of the phase-space that corresponds to the macrostate of entropy $S$. 

The physical processes in any thermodynamical system are irreversible. Consequently, any system always tends to its state of maximum entropy.

If we consider the maximal system, the universe, the situation is not so clear. In the past, the universe was hotter and at some point matter and radiation were in thermal equilibrium (i.e. in a state of maximum entropy); then, how can entropy still increase if it was at a maximum at some past time?

Some attemps were made \cite{Gold} to answer this question, invoking the expansion of the universe: the universe actually began in a state of maximum entropy, but because of the expansion, it was still possible for the entropy to continue growing.

The main problem with this type of explanation is that the effects of gravity cannot be neglected in the early universe \cite{p2}. Roger Penrose suggested that entropy might be  assigned  to the gravitational field itself. Though locally matter and radiation were in thermal equilibrium, the gravitational field should have been quite far from equilibrium, since gravity is an atractive force and the universe was initially structureless \cite{p2}. Consequently, the early universe was globally out of equilibrium, being the total entropy dominated by the entropy of the gravitational field. In absence of a theory of quantum gravity, a statistical measure of the entropy of the gravitational field is not possible. The study of the gravitational properties of macroscopic systems through classic general invariants, however, might be a suitable approach to the problem.

Penrose proposed that the Weyl curvature tensor can be used to specify the gravitational entropy. The Weyl tensor is a 4-rank tensor that contains the independent components of the Riemann tensor not captured by the Ricci tensor. It can be considered as the traceless part of the Riemann tensor.

The Weyl tensor can be obtained from the full curvature tensor by substracting out various traces. The Riemann tensor has 20 independent components, 10 of which are given by the Ricci tensor and the remaining by the Weyl tensor.  

The Weyl tensor is given by:
\begin{eqnarray}
C_{\alpha\beta\gamma\delta}&=&R_{\alpha\beta\gamma\delta}+\frac{2}{n-2}(g_{\alpha[\gamma}R_{\delta]\beta}-g_{\beta[\gamma}R_{\delta]\alpha})+\\ \nonumber
& + & \frac{2}{(n-1)(n-2)} R~g_{\alpha[\gamma}g_{\delta]\beta},\\
\end{eqnarray}
where $R_{\alpha\beta\gamma\delta}$ is the Riemann tensor, $R_{\alpha\beta}$ is the Ricci tensor, $R$ is the Ricci scalar, $[\;]$ refers to the antisymmetric part, and $n$ is the number of dimensions of space-time.

%In 4 dimensions the Weyl tensor is:

%\begin{eqnarray}
	%C_{\alpha\beta\gamma\delta}&=&R_{\alpha\beta\gamma\delta}+\frac{1}{2}\left(g_{\alpha\delta}R_{\gamma\beta}+g_{\beta\gamma}R_{\delta\alpha}-g_{\alpha\gamma}R_{\delta\beta}-g_{\beta\delta}R_{\gamma\alpha}\right)+\nonumber \\ &&+\frac{1}{6}\left(g_{\alpha\gamma}g_{\delta\beta}-g_{\alpha\delta}g_{\gamma\beta}\right)R.
%\end{eqnarray}

The behaviour of the Weyl tensor follows what is expected for the gravitational entropy throughout the history of the universe: it is zero in the (homogeneous) Friedmann-Robertson-Walker model and it is large in the Schwarzschild space-time.

Rudjord, Gr$\varnothing$n and Sigbj$\varnothing$rn \cite{Gron} made a recent attempt to develop a description of the gravitational entropy based on the construction of a scalar derived from the contraction of the Weyl tensor and the Riemann tensor. Their phenomenological approach is based on matching their description of the entropy of a black hole with the Hawking-Bekenstein entropy \cite{bek}. In particular, they calculated the entropy of Schwarzschild black holes and the Schwarzschild-de-Sitter space-time.

The main goal of the present study is to extend the calculation of the gravitational entropy to other space-time geometries and to analyse whether the proposal still works in more complex objects.
 
\section{Black hole entropy}
\label{sec:1}
In this section we provide a brief review of the work done by Rudjord et. al. They suggested an estimator for the gravitational entropy taking into account the Hawking-Bekenstein entropy. The entropy of a black hole can be described by the surface integral:
\begin{equation}\label{Sint}
S_{\rm{\sigma}} = k_{\rm{s}} \int_\sigma \vec \Psi \cdot \vec{d\sigma},
\end{equation}
where $\sigma$ is the surface of the horizon of the black hole and the vector field $\vec{\Psi}$ is:
\begin{equation}
\vec \Psi = P \vec e_{{\rm{r}}},
\end{equation}
with $\vec e_{\rm{r}}$ a unitary radial vector. The scalar $P$ is defined in terms of the Weyl scalar (W) and the Krestchmann scalar (R). It takes the form:
\begin{equation}\label{p1}
P^{2} = \frac{W}{R} = \dfrac{C^{\alpha\beta\gamma\delta}C_{\alpha\beta\gamma\delta}}{R^{\alpha\beta\gamma\delta}R_{\alpha\beta\gamma\delta}}\;.
\end{equation}
In order to find an acceptable description of the entropy of a black hole, it is required that Equation \ref{Sint} be equal at the horizon to the Hawking-Bekenstein entropy:
\begin{equation}
S_{\rm{\sigma}}=S_{\rm{HB}}.
\end{equation} 
The latter equation allows to calculate the constant $k_{\rm{s}}$:
\begin{equation}
k_{\rm{s}}=\frac{k_{\rm{B}}}{4l_{\rm{p}}^{2}}=\frac{k_{\rm{B}}c^{3}}{4G\hbar},
\end{equation}
where $k_{\rm{B}}$ is Boltzmann's constant, $l_{\rm{p}}^{2}= G\hbar c^{-3}$ is the Planck area, $G=6.674 \:\: . \:\: 10^{-8}$ $\rm{cm}^{-3}$$\rm{g}^{-1}$ $\rm{s}^{-2}$ is the constant of gravitation, and $\hbar=h /2 \pi$. As usual, $h$ is the Planck's constant. In what follows we will set the constant $k_{\rm{s}}$ equal to 1 since it plays the role of a scale factor in Equation \ref{Sint}. This simplifies the notation.

The entropy density can be then determined by means of Gauss's divergence theorem, rewriting Equation (\ref{Sint}) as a volume integral:
\begin{equation}\label{den}
\mathfrak{s} = k_{\rm{s}}\mid{ \nabla \cdot \vec \Psi}\mid.
\end{equation}
Here, the absolute value brackets were added to avoid negative or complex values of entropy.

Rudjord et. al. calculated the entropy density in Schwarzschild, de Sitter, and Schwarzschild-de-Sitter (SdS) space-times. Their investigations of the de Sitter space-time led them to conclude that the entropy of the cosmological horizon can not be gravitational. They suggested two different interpretations: either the horizon entropy is in general of different nature from gravitational entropy or, if a cosmological constant is introduced, there is a thermodynamical factor related to it. The Schwarzschild space-time seems to have only gravitational entropy. Then, the Schwarzschild and de Sitter space-times were interpreted as two extreme cases of the SdS space-time. This result is what is expected from a reasonable description of the gravitational entropy: large thermodynamical entropy in the early universe and large gravitational entropy around black holes.

In the next section we present the results of our calculations of the gravitational entropy and density for more complex objects: Reissner-Nordström, Kerr, Kerr-Newman black holes, and a special case of wormhole using Rudjord et. al.'s conjecture.

\section{Spherically symmetric space-times}
\subsection{Reissner-Nordström black holes}
The Reissner-Nordström metric \cite{Nord} is a spherically symmetric solution of the Einstein field equations \cite{Einstein}. It is not a vacuum-solution since the source has electric charge $Q$, and hence there is an electromagnetic field. The energy-momentum tensor of this field is:
\begin{equation}\label{e}
T_{\mu\nu}=-\dfrac{1}{\mu_{0}}(F_{\mu\rho}F_{\nu\rho}-\dfrac{1}{4}
g_{\mu\nu} F_{\rho\sigma}F^{\rho\sigma}),
\end{equation} 
where $F_{\rm{\mu\nu}}=\partial_{\rm{\mu}}A_{\rm{\nu}}-\partial_{\rm{\nu}}A_{\rm{\mu}}$ is the electromagnetic field strength tensor and $A_{\rm{\mu}}$ is the electromagnetic 4-potential. Outside the charged object the 4-current $j^{\rm{\mu}}$ is zero, so the Maxwell equations are:
\begin{eqnarray}
F^{\mu\nu};_{\mu}=0,\\
F_{\mu\nu};_{\sigma}+F_{\sigma\mu};{\nu}+F_{\nu\sigma};_{\mu}=0.
\end{eqnarray}

The solution for the metric is given by:
\begin{eqnarray}
ds^{2} &=& g_{\mu\nu}dx^{\mu}dx^{\nu} =  -\left(1-\dfrac{R_{s}}{r} + \dfrac{q^{2}}{r^{2}}  \right)dt^{2}+  \nonumber\\
&+&\dfrac{1}{\left(1-\dfrac{R_{s}}{r}+\dfrac{q^{2}}{r^{2}}\right)}dr^{2}+ r^{2}d\theta ^{2}+r^{2}\sin ^{2}\theta d\phi ^{2} ,\nonumber\\
\end{eqnarray}
where $R_{\rm{s}}=2GM/c^{2}$ is the Schwarzschild radius and:
\begin{equation}
q^{2}=\dfrac{GQ^{2}}{4\pi\varepsilon_{0} c^{4}},
\end{equation}
is related to the electric charge $Q$.

The study of the metric shows the presence of two different horizons: the outer ($r_{+}$) and the inner ($r_{-}$) horizons:
\begin{equation}\label{rnh}
r_{\pm}=M\pm\sqrt{M^{2}-Q^{2}}.
\end{equation}
If $Q\rightarrow 0 $, the outer horizon becomes Schwarzschild's and 
the inner horizon collapses into a central singularity.
We shall discuss a formulation of the gravitational entropy in terms of both horizons.

In order to find $P$, we calculate the Weyl scalar and the Kretschmann scalar. The results are:
\begin{eqnarray}
R & = & R^{\alpha\beta\gamma\delta}R_{\alpha\beta\gamma\delta}  \nonumber\\
& = & \dfrac{8\left( 6M^{2}r^{2}-12MrQ^{2}+7Q^{4} \right)}{r^{8}}
\end{eqnarray}
and
\begin{eqnarray}
W & = & C^{\alpha\beta\gamma\delta}C_{\alpha\beta\gamma\delta} \nonumber\\
& = & \dfrac{48}{r^{8}}(Q^{2}-Mr)^{2}.
\end{eqnarray}

Therefore, $P$ is given by:
\begin{eqnarray}
P^{2} & = & \dfrac{C^{\alpha\beta\gamma\delta}C_{\alpha\beta\gamma\delta}}{R^{\alpha\beta\gamma\delta}R_{\alpha\beta\gamma\delta}}\nonumber\\
& = & \dfrac{6M^{2}r^{2}-12MrQ^{2}+6Q^{4}}{6M^{2}r^{2}-12MrQ^{2}+7Q^{4}},\\
P & = & \sqrt{\dfrac{6M^{2}r^{2}-12MrQ^{2}+6Q^{4}}{6M^{2}r^{2}-12MrQ^{2}+7Q^{4}}}.
\end{eqnarray}

According to Rudjord et. al. \cite{Gron}, in order to find the gravitational entropy we must operate in a 3-space. Therefore, the spatial metric is defined as:
\begin{equation}
h_{ij}= g_{ij}-\frac{g_{i0}g_{j0}}{g_{00}}\;,
\label{3DST}
\end{equation}
where $g_{\rm{\mu\nu}}$ is the 4-dimensional space-time metric and the Latin indices denote spatial components, $i$, $j =$1,2,3. Then, the infinitesimal surface element is given by:
\begin{equation}
d \sigma=\frac{\sqrt{h}}{\sqrt {h_{rr}}} d\theta d\phi \;.
\label{infele}
\end{equation}

From Equations \ref{3DST}, and \ref{infele} in the Reissner-\\
Nordtröm space-time, the space metric and the infinitesimal surface element are:
\begin{equation}
h_{ij} = \mathrm{diag}\left[\left(1-\dfrac{R_{s}}{r}+\dfrac{Q^{2}}{r^{2}}  
\right)^{-1}, r^2, r^2\sin\theta\right],
\end{equation}
\begin{eqnarray}
d \sigma = r^2 \sin\theta d\theta d\phi.
\end{eqnarray}

The integration over the two different horizons must be done carefully since there is a singularity at the origin. The simple method that Rudjord et. al. developed in the case of the Schwarzschild black hole was to integrate over a small sphere with radius $\epsilon$ around the origin, and to substract this from integral \ref{Sint}. This yields:
\begin{eqnarray}
S_{\pm} &=& k_{\rm{RN}} \int_0^\pi \int_0^{2\pi}  P(r_{\pm})r_{\pm}^2 \sin \theta 
d \theta d \phi \nonumber\\
&-& k_{\rm{RN}}\int_0^\pi \int_0^{2\pi}  P(\varepsilon)\varepsilon^2\sin \theta 
d \theta d \phi .
\end{eqnarray}
The scalar $P$ does not depend on $\theta $ and $\phi $, so we have:
\begin{eqnarray}
S_{\pm} &=& k_{\rm{RN}} \left[ P(r_{\pm})r_{\pm}^2 - P(\varepsilon)\varepsilon^2
\right]\int_0^{2\pi} \int_0^\pi \sin \theta d \theta d \phi \notag. \\
\end{eqnarray}
When $\epsilon \rightarrow 0$, the entropy of the horizons is obtained:
\begin{equation}\label{s+}
S_{\pm}=k_{\rm{RN}}4\pi r^{2}{\pm}\sqrt{  \dfrac{6M^{2}r_{\pm}^{2}-12Mr_{\pm}Q^{2}+6Q^{4}}{6M^{2}r_{\pm}^{2}
-12Mr_{\pm}Q^{2}+7Q^{4}}}.
\end{equation}
%and at the inner horizon:
%\begin{equation}
%S_{-}=k_{\rm{RN}}4\pi R^{2}_{-}\sqrt{\dfrac{6M^{2}R_{-}^{2}-12MR_{-}Q^{2}+6Q^{4}}{6M^{2}R_{-}^{2}
%-12MR_{-}Q^{2}+7Q^{4}}}.
%\end{equation}
By replacing Equation \ref{rnh} into Equation \ref{s+} we can see that when $Q\rightarrow 0 $, the entropy of the outer horizon becomes 
the Schwarzschild entropy \cite{Gron}. For the inner horizon we obtain an indeterminate form:
\begin{equation}
\lim_{Q\rightarrow 0} S_{-} \longrightarrow 0 \times \dfrac{0}{0}. 
\end{equation}

A possible interpretation of the last limit is the following: when the electric charge goes to zero, the inner horizon collapses into the singularity. In other words, the entropy of the inner horizon goes to the entropy of the singularity. General Relativity is not an epistemologically well-determined theory: the language of the theory says nothing about the singularity because continuum space-time does not exist there. 

We now calculate the entropy density using Equation \ref{den}. We get:
\begin{eqnarray}
\mathfrak{s} & = & k_{RN} \left| \dfrac{1}{r^{2}}\sqrt{\left(1-\dfrac{R_{s}}{r}+\dfrac{Q^{2}}{r^{2}}  
\right)} \dfrac{\partial }{\partial r}(r^{2}  P) \right| \\ \nonumber
 & = &  k_{RN} \left| \dfrac{1}{r^{2}} \sqrt{\left(1-\dfrac{R_{s}}{r}+\dfrac{Q^{2}}{r^{2}}  \right)} \times \left(s_{\rm{RN1}} + s_{\rm{RN2}}\right) \right|,
\end{eqnarray} 
where:
\begin{equation}
s_{\rm{RN1}}= 2\,r\sqrt {{\frac {6\,{r}^{2}-12\,r{Q}^{2}+6\,{Q}^{4}}{6\,{r}^{2}-12\,r{Q}^{2}+7\,{Q}^{4}}}},
\end{equation}
and
\begin{equation}
s_{\rm{RN2}}=  \dfrac{1}{2} \,{r}^{2} \left[\dfrac{\sqrt{6}\,Q^{4}\,M}{\left({6\,{r}^{2}-12\,r{Q}^{2}+7\ Q^{4} }\right)^{\frac{3}{2}}} \right].
\end{equation}

\begin{figure}[h]
\includegraphics[height=6cm, width=9cm]{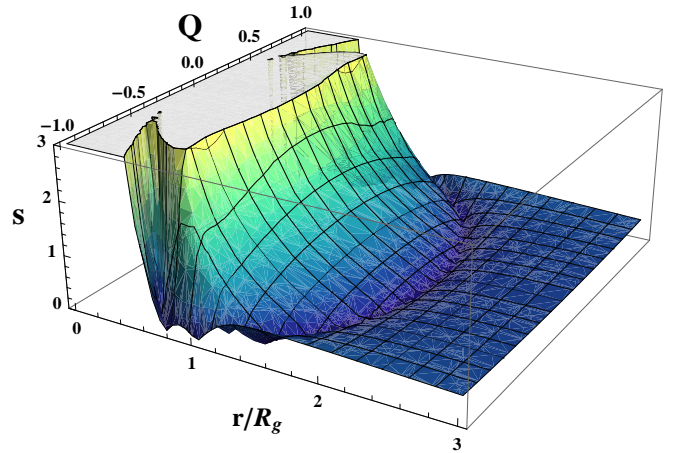}
\caption{  Entropy density of a Reissner-Nordström black hole. Here, $c= G =k_{RN} = 1$, $M=1$. }
\label{fig:RN_entropy_density}
\end{figure}

Again, if $Q\rightarrow 0 $, we obtain the same result as in the Schwarzschild case. 
Figure \ref{fig:RN_entropy_density} shows a plot of $\mathfrak{s}$ for different values of the charge.

\begin{figure}[h]
      \includegraphics[height=5cm, width=7cm]{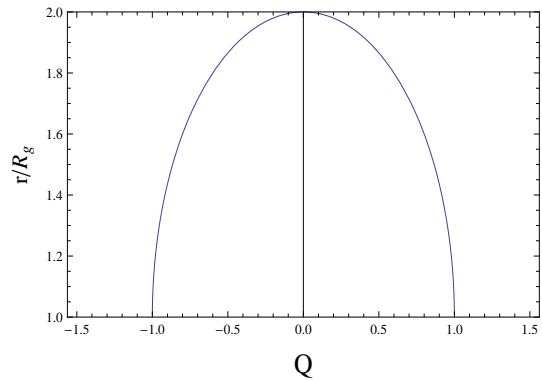}
       \caption{  Radii of the outer horizon. Here, $c= G = 1$, $M=1$. }
        \label{radiusRN}
 \end{figure}

Figure \ref{radiusRN} shows the value of the radius of the outer horizon for a given range of the charge. The radius is only well defined for $Q \in[-1,+1]$. Outside this interval the horizon does not exist and the singularity is naked. According to the cosmic censorship conjecture \cite{p4}, singularities only occur if they are hidden by an event horizon. If this conjecture is valid, the value of the charge has to be restricted.
 
\begin{figure}[h]
      \includegraphics[height=6cm, width=8cm]{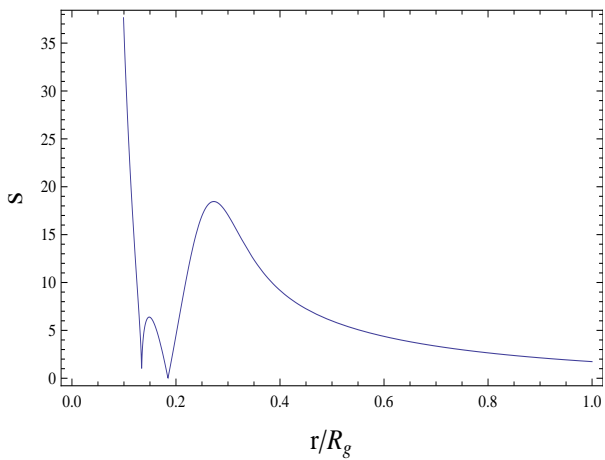}
       \caption{Entropy density of a Reissner-Nordström black hole for $Q=0.5$. Here, $c= G =k_{RN} = 1$, $M=1$. }
       \label{2drn}
 \end{figure}

Figure \ref{2drn} shows a plot of the entropy density for a fixed value of the charge. The function $\mathfrak{s}$ is not defined at $r=0$ and it tends asymptotically to zero for large values of the radius. This is actually what it is expected of a good description of the entropy density for Reissner-Nordström black holes: $\mathfrak{s}$ can not be defined at the origin because of the presence of a singularity, and goes to zero as the field decreases.
   
Furthermore, one of the minima of the function $\mathfrak{s}$ is at $r = 0.13$ where the inner Cauchy horizon is located. This horizon is near two relative maxima and an other relative minimum of the function $\mathfrak{s}$. A possible explanation is that in the case of Reissner-Nordström black hole, near the Cauchy horizon, there is a small region of space-time where the curvature is high and potentially unstable against external perturbations, as shown by Poisson and Israel \cite{po-is}. The behaviour of the entropy density for $r \in [0.13,0.3]$ could then be reflecting the accumulation of gravitational energy and entropy in that region, and the instability of the inner horizon. 
%Figure \ref{RN11} showns the plot of the entropy density for the permitted values of the charge.

\subsection{Wormholes}
\label{secw}
\subsubsection{General features}

A wormhole is a region of space-time with non-trivial topology. It has two mouths connected by a throat. The mouths are not hidden by event horizons as in the case of black holes, and, in addition, there is no singularity to avoid the passage of particles from one side to the other. Contrary to black holes, wormholes are holes in space-time, i.e., their existence implies a multiple-connected space-time \cite{Gus}.

There are many types of wormhole-like solutions of the Einstein field equations \cite{Visser}. Let us consider the static spherically symmetric line element:
\begin{equation}
ds^{2} = -\exp^{2\Phi}c^{2}dt^{2}+\frac{dr^{2}}{1-\frac{b}{r}}+r^{2}\left(d\theta^{2}+{\sin{\theta}}^{2}d\phi^{2}\right).
\end{equation}
Here, $\Phi=\Phi(r)$ and $b=b(r)$ are two arbitrary functions to be constrained by the required properties of the wormhole. The function $b=b(r)$ is called ``the shaped function'' because it determines the spatial shape of the wormhole, whereas $\Phi(r)$ determines the gravitational redshift and is called the ``redshift function''. The radial coordinate $r$ has a special significance: $2\pi r$ is the circumference of a circle centered on the wormhole's throat. It decreases from $+\infty$ to a minimum value $b_{0}$, the radius of the throat; then it increases from $b_{0}$ back to $+\infty$ as distance to the throat increases. 

In order to have a wormhole which is transversable in principle, we demand \cite{Morris} \cite{Gus}:
\begin{enumerate}
\item The function $\Phi(r)$ must be finite everywhere (to be consistent with the absence of event horizons).
\item In order for the spatial geometry to tend to an appropiate asymptotically flat limit, it must happen that:
\begin{displaymath}
\lim_{r\to \infty} \frac{b(r)}{r} \rightarrow 0,
\end{displaymath}
and,
\begin{displaymath}
\lim_{r\to \infty} \Phi(r) \rightarrow 0.
\end{displaymath}
\end{enumerate}

We can impose certain restrictions on the stress-energy that generates the wormhole's curvature. These ``junction conditions'' allow to determine the specific form of the functions $b(r)$ and $\Phi(r)$.

One possible case is to confine the exotic material, that keeps the throat open, to the interior of a sphere of surface radius $r=R_{\rm{s}}$. The vacuum region outside $r=R_{\rm{s}}$ constrains the external space-time geometry to have the standard Schwarzschild form \cite{Morris}:
\begin{displaymath}
\begin{array}{ll}
b(r)  =  b(R_{\rm{s}})  =  B = \textrm{const} & \textrm{at\ \ $r\ge R_{\rm{s}}$},\\
\Phi(r)  =  \frac{1}{2}\ln{\left(1-\frac{B}{r}\right)} &  \textrm{at\ \ $r\ge R_{\rm{s}}$}.
\end{array}
\end{displaymath}

There are some other solutions in which the exotic material is limited to the throat vecinity. Morris and Thorne \cite{Morris} gave an example of a wormhole whose interior solution in and around the throat has exotic material and it is joined onto non-exotic matter at a radius $r_{\rm{c}}$. There is a surface layer at $r=R_{\rm{s}}$ of thickness $\Delta R$ where the tension $\tau$ drops to zero. Finally, the external region has the Schwarzschild form for $r \geq R+\Delta R$.
  
The complete wormhole solution reads:
\begin{displaymath}
b(r)= \left\{ \begin{array}{ll}
\left(b_{0}r\right)^{\frac{1}{2}} & \textrm{at $b_{\rm{0}}\leq r \leq r_{\rm{c}}$},\\
1/100r & \textrm{at $r_{\rm{c}}\leq r \leq R_{\rm{s}}$},\\
\frac{1}{3}\left[\left(r^{3}-R_{\rm{s}}^{3}\right)/R_{\rm{s}}^{2}\right]+R_{s}/100 & \textrm{at $R_{\rm{s}}\leq r \leq R_{{\rm l}}$},\\
B\equiv R_{\rm{s}}/100 & \textrm{at $R_{\rm{l}} \leq r$},
\end{array}\right.
\end{displaymath}

\begin{displaymath}
\Phi(r)= \left\{ \begin{array}{ll}
\Phi_{0}\cong -0.01 & \textrm{at $b_{\rm{0}} \leq r \leq R_{{\rm l}}$},\\
\frac{1}{2}\ln{\left(1-\frac{B}{r}\right)} & \textrm{at $R_{{\rm l}} \leq r$},
\end{array}\right.
\end{displaymath}
where $r_{\rm{c}}=10^{4}b_{\rm{0}}$, $R_{{\rm l}} = R_{\rm{s}}+\Delta R$ , and $\Delta R=1/100 R$.

\subsubsection{Entropy density}

Since one of the main features of wormholes is the absence of event horizons, the formulation of the entropy in terms of a surface integral on the event horizon is meaningless. As we shall see, the calculation of the entropy density using the Rudjord et. al.'s conjecture, however, is still possible.

We begin the calculus in the region closer to the throat and with exotic matter. The metric takes the form:
\begin{equation}\label{wm}
ds^{2} = -\exp^{2\Phi_{0}}c^{2}dt^{2}+\frac{dr^{2}}{1-\sqrt{\frac{b_{0}}{r}}}+r^{2}\left(d\theta^{2}+{\sin{\theta}}^{2}d\phi^{2}\right),
\end{equation}
where $\Phi_{0}\cong -0.01$ and $b_{\rm{0}}$ is the radius at the throat.

The calculation of the Weyl and Kretschmann scalars gives:
\begin{eqnarray}
W & = & C^{\alpha\beta\gamma\delta}C_{\alpha\beta\gamma\delta} \nonumber\\
& = & \frac{25 b_{0}}{12 r^{5}},
\end{eqnarray}
and
\begin{eqnarray}
R & = & R^{\alpha\beta\gamma\delta}R_{\alpha\beta\gamma\delta}  \nonumber\\
& = & \frac{9b_{0}}{2r^{5}},
\end{eqnarray}
respectively.

The scalar $P$ takes the simple form:
\begin{equation}
P= \frac{5}{3\sqrt{6}}
\end{equation}

From Equation (\ref{wm}) the spatial metric is:
\begin{equation}
h_{ij} = \mathrm{diag}\left(\frac{1}{1-\sqrt{\frac{b_{0}}{r}}}  , r^{2}, r^{2}\sin\theta^{2}\right).
\end{equation}
In terms of $P$ and the covariant derivative the entropy density is:
\begin{equation}
\mathfrak{s}= k_{{\rm w}}\left|\sqrt{\frac{2}{27}} 5 \sqrt{\left[1-\sqrt{\frac{b_{0}}{r}}\right]} \frac{1}{r}\right|.
\end{equation}

\begin{figure}[h]
\includegraphics[height=6cm, width=8cm]{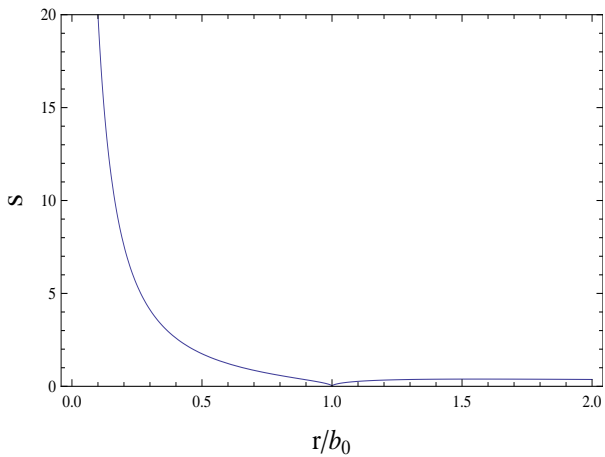}
\caption{  Entropy density for a wormhole with exotic matter limited to the throat vecinity. Here $k_{\rm{w}}=1$.}
\label{fig:wormhole}
\end{figure}

The plot of $\mathfrak{s}$ as a function of $r$ is shown in Figure \ref{fig:wormhole}.

We observe in Figure \ref{fig:wormhole} that the entropy density is zero at the throat (where the exotic matter is confined) and then increases smoothly to then to tend asymptotically to zero at large distances from the object. For $r\leq b_{0}$ the solution obtained has no physical meaning because the wormhole metric is defined for $r\in [b_{0},+\infty)$. This result is what we expect, interpreting the wormhole as a surgery construction out from two symmetric Schwarzschild space-times.

\section{Axisymmetric space-times}

\subsection{Kerr black holes}

The solution of the field equations for a rotating body of mass $M$ and angular momentum per unit mass $a$ was found by Kerr \cite{Kerr}. Since the Boyer-Lindquist coordinates present coordinate singularities in the event horizons, we work in the Kerr coordinates where these pathologies have been removed. The Boyer-Lindquist and the Kerr coordinates are related by the following coordinate transformation:
\begin{eqnarray}
dv & = & dt+\dfrac{(r^2-a^2)}{\Delta}dr\\
d\chi & = & d\phi + \dfrac{a}{\Delta}dr
\end{eqnarray}

The metric can be written in matrix form as:
%\begin{multicols}{1}
%\lipsum[1-3]
%\end{multicols}
%\par\noindent\rule{\dimexpr(0.5\textwidth-0.5\columnsep-0.4pt)}{0.4pt}%
%\rule{0.4pt}{6pt}
%\begin{widetext}
%g_{\mu\nu}=
%\begin{array}[pos]{spalten}
%-\dfrac{\Delta-a^2\sin^2\theta}{\rho^2}  & 1 & 0 & \dfrac{-a\sin^2\theta(r^2+a^2-\Delta)}{\rho^2} \\ 
%1 & 0 & 0 & -a\sin^2\theta \\ 
%0 & 0 & \rho^2 & 0 \\ 
%\dfrac{-a\sin^2\theta(r^2+a^2-\Delta)}{\rho^2} & -a\sin^2\theta & 0 &  \dfrac{\left[(r^2+a^2)^2-a^2\Delta\sin^2\theta\right]}{\rho^2}\sin^2\theta  
 
%\end{array} .
\begin{eqnarray}
g_{\mu\nu}=\begin{pmatrix}
\alpha & 1 & 0 & \beta\\
1 & 0 & 0 & -a\sin^2\theta \\ 
0 & 0 & \rho^2 & 0 \\ 
\beta & -a\sin^2\theta & 0 &  \delta
\end{pmatrix},
\end{eqnarray}
where:
\begin{eqnarray}
\alpha &=& -\dfrac{\Delta-a^2\sin^2\theta}{\rho^2},\\
\beta & = & \dfrac{-a\sin^2\theta(r^2+a^2-\Delta)}{\rho^2},\\
\delta & = &  \dfrac{\left[(r^2+a^2)^2-a^2\Delta\sin^2\theta\right]}{\rho^2}\sin^2\theta,
\end{eqnarray}
%$\bigl( \begin{smallmatrix}
%-\dfrac{\Delta-a^2\sin^2\theta}{\rho^2}  & 1 & 0 & \dfrac{-a\sin^2\theta(r^2+a^2-\Delta)}{\rho^2} \\ 
%1 & 0 & 0 & -a\sin^2\theta \\ 
%0 & 0 & \rho^2 & 0 \\ 
%\dfrac{-a\sin^2\theta(r^2+a^2-\Delta)}{\rho^2} & -a\sin^2\theta & 0 &  \dfrac{\left[(r^2+a^2)^2-a^2\Delta\sin^2\theta\right]}{\rho^2}\sin^2\theta   
%\end{smallmatrix} \bigr)$

%\begin{eqnarray}
%g_{\mu\nu}=\begin{pmatrix}
%-\left(1-\dfrac{2Mr-Q^2}{\rho^2}\right) & 1 & 0 & -a\dfrac{(2Mr-Q^2)}{\rho^2}\sin^2\theta \\ 
%1 & 0 & 0 & -a\sin^2\theta \\ 
%0 & 0 & \rho^2  & 0 \\ 
%-a\dfrac{(2Mr-Q^2)}{\rho^2}\sin^2\theta & -a\sin^2\theta & 0 & \dfrac{\left[(r^2+a^2)^2-\Delta_{\textit{\tiny{\rm{KN}}}} a^2 \sin^2\theta\right]}{\rho^2}\sin^2\theta
%\end{pmatrix} .
%\end{eqnarray}

and:
\begin{eqnarray*}
\Delta & = & r^{2}+a^{2}-2Mr, \\
\rho^{2} & = & r^{2} + a^{2}\cos^{2}\theta,\\
a & = & \dfrac{J}{M}. 
\end{eqnarray*}
Here $J$ represents the angular momentum of the black hole and the inner and outer horizons occur at:
\begin{equation}
r_{\pm} = M \pm \sqrt{M^2 - a^2}.
\end{equation}

\begin{figure}[h]
      \includegraphics[height=5cm, width=7cm]{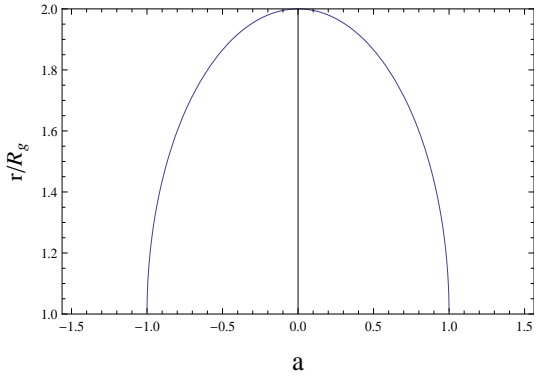}
       \caption{Radii of the outer horizon in the case of Kerr black hole. Here, $c= G = 1$, $M=1$.}
       \label{radiuskerr}
 \end{figure}

Figure \ref{radiuskerr} shows a plot of the radius of the outer horizon as a function of the angular momentum. Again, assuming the cosmic censorship conjecture, the values of the angular momentum are restricted to $a \in [-1,+1]$.

Just as the Schwarzschild solution is the unique vacuum solution of Einstein's field equations (as result of Israel's theorem), the Kerr metric is the unique stationary axisymmetric vacuum solution (Carter-Robinson theorem). The calculation of the scalars yields:
\begin{eqnarray}
R_{\mu\nu} &=& 0,\\
C^{\alpha\beta\gamma\delta}C_{\alpha\beta\gamma\delta} & = & R^{\alpha\beta\gamma\delta}R_{\alpha\beta\gamma\delta},\\
P^{2} & = & 1,\\
P & = & +1\;.\label{p}
\end{eqnarray}

Since $P = 1$, the result of the calculation of the entropy in the outer event horizon is equal to the area of this horizon. The metric in this horizon takes the form:
\begin{eqnarray}\label{sup}
ds^{2}  = \dfrac{\left[ (r_+^2+a^2)^2-\Delta a^2\sin^2\theta \right]}{\rho^2}\sin^2\theta d\chi^2+\rho^2d\theta^2.
\end{eqnarray}
The $\Delta$ function can be written as $\Delta = (r-r_-)(r-r_+) $, where $r_{-}$ and $r_{+}$ are the radii of the inner and outer horizon respectively. From Equation \ref{sup}, replacing $\Delta(r=r_+)=0$, we obtain:
\begin{equation}
ds^{2}  = \dfrac{ (r_+^2+a^2)^2}{\rho^2}\sin^2\theta d\chi^2+\rho^2d\theta^2.
\end{equation}

Finally, the calculation of the area gives:
\begin{equation}
A_+=\iint \sqrt{g_{\theta\theta}g_{\chi\chi}}d\theta d\chi,
\end{equation}
\begin{eqnarray}
A_+ &=&\iint \sqrt{ \dfrac{(r_+^2+a^2)^2}{\rho^2}    \sin^2\theta\rho^2    }d\theta d\chi \\ \nonumber
& = & \iint (r_+^2+a^2)\sin\theta d\theta d\chi.
\end{eqnarray}
\begin{eqnarray}
A_+&=&(r_+^2+a^2)\int_0^{2\pi}d\chi \int_0^{\pi}\sin\theta d\theta ,\nonumber\\
A_+&=& 4\pi(r_+^2+a^2).
\end{eqnarray}

In the same way we can make the calculation for $r = r_{-}$:
\begin{equation}
A_-= 4\pi(r_-^2+a^2).
\end{equation}
The entropy for both horizons takes the simple form:
\begin{eqnarray}
S_+&=&k_{\rm{K}} 4\pi(r_+^2+a^2),\nonumber\\
S_-&=&k_{\rm{K}} 4\pi(r_-^2+a^2).
\end{eqnarray}
For $a \rightarrow 0$, the correct Schwarzschild limit is obtained.

In the case of axisymmetric space-times, it is not possible to calculate the spatial metric (see Equation \ref{3DST}) because of the presence of the metric coefficient $g_{t \phi}$: as the body is rotating the spatial position of each event in the Kerr space-time depends on time. The covariant divergence, then, is obtained from the determinant of the full metric:
\begin{equation}\label{for}
\nabla \cdot \vec \Psi = \dfrac{1}{\sqrt{-g}} \left(\dfrac{\partial}{\partial r} \sqrt{-g}\;P\right),
\end{equation}
where $g$ is the determinant of the metric and is given by:
\begin{equation}\label{g}
g=-\dfrac{1}{4}\left[a^2 + 2 r^2 + a^2 \cos2 \theta \right]^2 \sin^2\theta.
\end{equation}

By introducing Equation \ref{g} into \ref{for}, we calculate the entropy density. The result is:
\begin{equation}\label{densitykerr}
\mathfrak{s}=k_{\rm{K}} \left| \dfrac{4 r}{a^2 + 2 r^2 + a^2 \cos2 \theta}\right|.
\end{equation}
We can see that the entropy density does not depend explicitly on the mass of the black hole.

\begin{figure}[h!]
\begin{center}
  \includegraphics[height=5cm, width=9cm,]{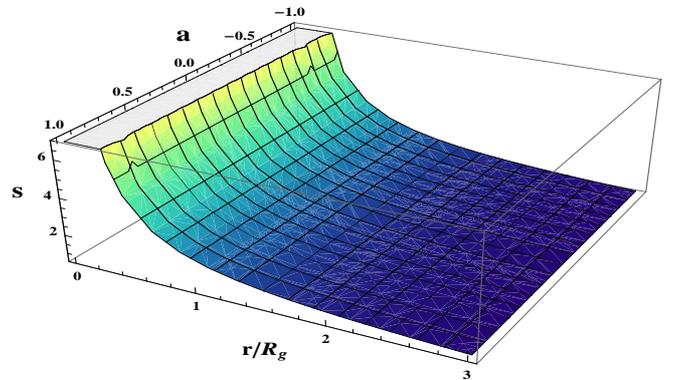}
\end{center}
\caption{Entropy density of a Kerr black hole as a function of the radial coordinate and the angular momentum. Here, $c= G =k_{K} = 1$, and $\theta = \pi/2$.}
\label{densitykerr1}
\end{figure}

A 3-dimensional plot of Equation \ref{densitykerr} is shown in Figure \ref{densitykerr1} for $\theta = \pi / 2$ and different values of the angular momentum. The entropy density is everywhere well defined except for $\theta=\pi/2$ and $r=0$, where the ring singularity is located. At large distances, $r\rightarrow \infty$, the function $\mathfrak{s}$ goes to zero, as expected.

%\begin{figure}[h!]
% \begin{center}
%  \includegraphics[height=5cm, width=9cm,]{sale.eps}
% \end{center}
%\caption{Entropy density of a Kerr black-hole as a function of the radius and the angular momentum. Here, $c= G =k_{K} = 1$, and $\theta=\pi/2$.}
%\label{sale}
%\end{figure}

\subsection{Kerr-Newman black hole}

The Kerr-Newman metric of a charged spinning black hole is the most general black hole solution. It was found by Newman et al. in 1965 \cite{Newman}.

As in the case of the Kerr black hole, in order to avoid the coordinate singularities in the event horizons, we make the following change of coordinates:

\begin{eqnarray}
dv & = & dt+\dfrac{(r^2-a^2)}{\Delta_{\textit{\tiny{\rm{KN}}}}}dr,\\
d\chi & = & d\phi + \dfrac{a}{\Delta_{\textit{\tiny{\rm{KN}}}}}dr,
\end{eqnarray}
where:
\begin{equation}
\Delta_{\textit{\tiny{\rm{KN}}}}  =  r^{2}+a^{2}+Q^{2}-2Mr.
\end{equation}

The metric takes the matrix form:
\begin{eqnarray}
g_{\mu\nu}=\begin{pmatrix}
\gamma & 1 & 0 & \eta \\ 
1 & 0 & 0 & -a\sin^2\theta \\ 
0 & 0 & \rho^2  & 0 \\ 
\eta & -a\sin^2\theta & 0 & \xi
\end{pmatrix} .
\end{eqnarray}
Here, $\rho^{2}_{\textit{\tiny{\rm{KN}}}}  =  r^{2} + a^{2}\cos^{2}\theta$, and:
\begin{eqnarray}
\gamma & = & -\left(1-\dfrac{2Mr-Q^2}{\rho^2}\right),\\
\eta & = & -a\dfrac{(2Mr-Q^2)}{\rho^2}\sin^2\theta,\\
\xi & = & \dfrac{\left[(r^2+a^2)^2-\Delta_{\textit{\tiny{\rm{KN}}}} a^2 \sin^2\theta\right]}{\rho^2}\sin^2\theta.
\end{eqnarray}

\begin{figure}[h]
      \includegraphics[height=6cm, width=8cm]{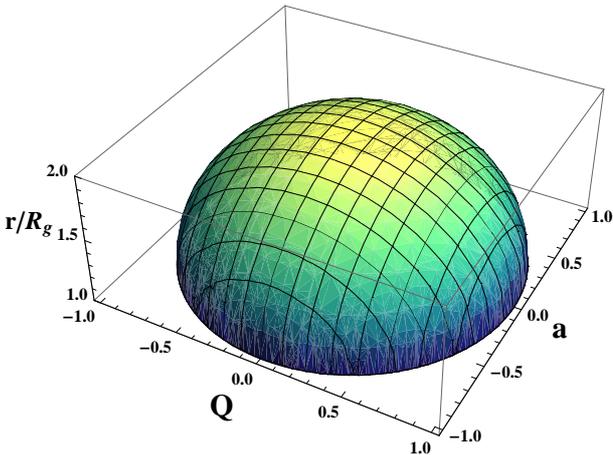}
       \caption{Radii of the outer horizon for the Kerr-Newman space-time. Here, $c= G = 1$, $M=1$. }
        \label{rayonKN}
 \end{figure}

The Kerr-Newman solution depends on three parameters: the mass, the angular momentum and the charge. We show, in Figure \ref{rayonKN}, that the radius of the outer horizon is well defined for $a \in[-1,1]$ and $Q \in[-1,1]$. This is clearly a combination of the previous cases.

The calculation of the determinant $g=\textsf{det}(g_{\mu\nu})$ gives:
\begin{equation}
g=-\sin^2\theta\left(r^4+2\cos^2\theta a^2r^2+\cos^4\theta a^4\right),
\end{equation}
which has the same form as in the Kerr space-time.

Since the Kerr-Newman space-time is a non-vacuum solution of Einstein's field equations because of the presence of an electromagnetic field, the Weyl and\\ Kretschmann scalars are not equal. The calculation of the scalar $P$ yields:
\begin{eqnarray}
P^2& =  & \dfrac{W}{R}\nonumber\\
P & = & \sqrt{\dfrac{W}{R}} = \sqrt{\frac{A}{B}},
\end{eqnarray}
where:
\begin{eqnarray}\label{A}
A & =  &-48\, ({M}^{2}{\xi}^{6}+10\,{\xi}^{4}{r}\,
{M}\,{Q}^{2}-{\xi}^{4}{Q}^{4}-15\,{\xi}^{4}{M}^{2}{r}^{2}\nonumber\\
&+& 6\,{\xi}^{2}{Q}^{4}{r}^{2}+15\,{r}^{4}
{M}^{2}{\xi}^{2}- 20\,{M}\,{Q}^{2}{r}^{3}{\xi}^{2}\nonumber\\
&-& {M}^{2}{r}^{6}+2\,{M}{r}^{5}{Q}^{2}-{Q}^{4}{r}^{4}), \nonumber\\
\end{eqnarray}

\begin{eqnarray}
B &= &-8\, (-6\,{M}^{2}{r}^{6}+90\,{\xi}^{2}
{r}^{4}{M}^{2}-90\,{\xi}^{4}{M}^{2}{r}^{2}\nonumber\\
&+& 6\,{\xi}^{6}{M}^{2}+12\,{r}^{5}{Q}^{2}M- 120\,{\xi}^{2}M{Q}^{2}{r}^{3}\nonumber\\
&+& 60\,M{\xi}^{4}{Q}^{2}r-7\,{
Q}^{4}{r}^{4}+34\,{\xi}^{2}{Q}^{4}{r}^{2}-7\,{\xi}^{4}{Q}^{4}),\nonumber\\
\label{B}
\end{eqnarray}
and, 
\begin{equation}
\xi = a \cos\theta  \;.
\label{xi}
\end{equation} 

We compute the entropy density using Equation \ref{for}. The function we find, however, is singular for certain values of the radius:
\begin{equation}
\mathfrak{s}=k_{\rm{KN}}\frac{1}{2}\left[\left(\frac{1}{g}\sqrt{A}g'+\frac{A'}{\sqrt{A}}\right)\frac{1}{\sqrt{B}}-\left(B'\sqrt{\frac{A}{B^{3}}}\right)\right],
\end{equation}
since $R$, the Kretschmann scalar, is a polynomial of order six with at least one positive real root. 
For example, if we solve the equation $B =0$ for $a= 0.5 $, $\theta= \pi/4$, $Q=0.6$, and $M =1$, the roots of the Kretschmann scalar are:
\begin{eqnarray}
r_{1} & = & -1.21\\
r_{2} & = & -0.27\\
r_{3} & = & 0.007\\
r_{4} & = & 0.22\\
r_{5} & = & 0.53\\
r_{6} & = & 1.45
\end{eqnarray}

Since the only difference between Kerr space-time and Kerr-Newman space-time is the $P$ scalar (the determinant $g$ is the same), we conclude that Rudjorn et. al.'s proposal for the entropy scalar (see Equation \ref{p1}) can not be used to calculate the entropy density in the Kerr-Newman space-time. Hence, we propose this alternative definition for $P$:
\begin{equation}\label{p2}
P=C^{\alpha\beta\gamma\delta}C_{\alpha\beta\gamma\delta}.
\end{equation}
For a Kerr-Newman black hole Equation \ref{p2} yields:
\begin{equation}
P=-\dfrac{48}{(r^2+a^2\cos^2\theta)^6}\left(p_{1}\;p_{2}\right)
\end{equation}
where:
\begin{eqnarray*}
p_{1}& = & -Mr^3-3Mr^2a \cos^2\theta + 3Mr \cos^2\theta a^2+\\
& + & M\cos^3\theta a^3 -\cos^2\theta a^2 Q^2+2\cos\theta a Q^2 r + Q^2 r^2,\\
\end{eqnarray*}
\begin{eqnarray*}
p_{2} & = & +Mr^3-3Mr^2 \cos\theta a -3M\cos^2\theta a^2 r +\\
& + & M \cos^3\theta a^3+\cos^2\theta a^2 Q^2+2\cos\theta a Q^2 r-Q^2 r^2\\
\end{eqnarray*}
We compute the entropy density using Equation \ref{for}. The result is:
\begin{equation}
\mathfrak{s}=\left| -\dfrac{96\:k_{\rm{KN}}}{(r^2+a^2\cos^2\theta)^7} \left( s_{1} + s_{2} - s_{3}+s_{4}\right)\right|,
\end{equation}
and:
\begin{eqnarray*}
s_{1} & = & 5 \cos^6\theta a^6 Q^2 M -20\cos^6\theta a^6 r M^2+90 \cos^4\theta M^2 r^3,\nonumber\\
s_{2} & = & 11\cos^4\theta a^4 Q^4 r-75 \cos^4\theta a^4 Q^2 m r^2-\nonumber\\
&& 26\cos^2\theta a^2 Q^4 r^3,\nonumber\\
s_{3} & = & 48\cos^2\theta a^2 r^5 M^2+75\cos^2\theta a^2 r^4 Q^2 M,\nonumber\\
s_{4} & = & 3r^5 Q^4 +2 r^7 M^2-5r^6Q^2M.\nonumber
\end{eqnarray*}

\begin{figure}[h!]
 \begin{center}
  \includegraphics[height=5cm, width=8cm,]{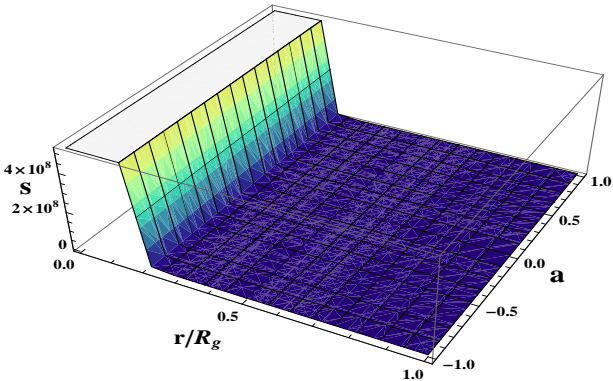}
 \end{center}
\caption{Entropy density of a Kerr-Newman black-hole as a function of the radial coordinate and the angular momentum. Here, $c= G =k_{K} = 1$,  $\theta=\pi/2$, and $Q = 0.6$.}
\label{KN13D}
\end{figure}

\begin{figure}[h!]
 \begin{center}
  \includegraphics[height=5cm, width=9cm,]{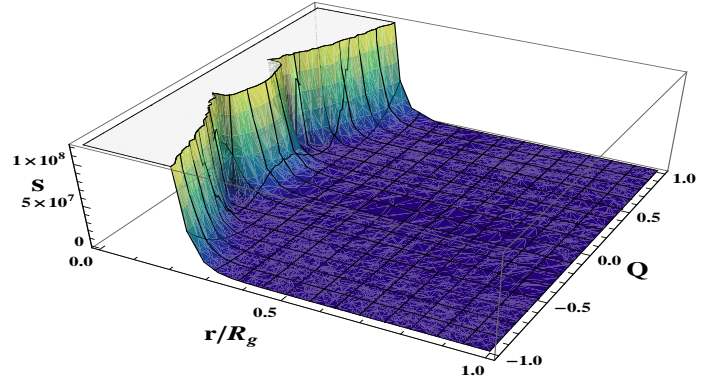}
 \end{center}
\caption{Entropy density of a Kerr-Newman black-hole as a function of the radial coordinate and the charge. Here, $c= G =k_{K} = 1$,  $\theta=\pi/2$, and $a = 0.6$.}
\label{KN23D}
\end{figure}

In Figures \ref{KN13D} and \ref{KN23D} it is shown the plots of the entropy density of Kerr-Newman black hole as a function of the radial coordinate and the angular momentum or the charge respectively. In both cases the entropy density diverges for $r=0$ and $\theta =\pi/2$ as  expected.

Figure \ref{tabla} shows a table with values of the entropy density at different $\theta$, in the core of the object and in both horizons. It can be seen that only in the equatorial plane the entropy density diverges, where the ring singularity is located. For $ \theta \neq \pi/2$, the entropy density is everywhere well defined.

\begin{figure}[h!]
\begin{center}
\begin{tabular}{|c|c|c|c|}
\hline $\theta$ & $ \lim_{r\rightarrow 0} \dfrac{\mathfrak{s}}{k_{\rm{KN}}} $ & $\lim_{r\rightarrow r_+} $ & $\lim_{r\rightarrow r_-}$ \\ 
\hline $\pi/2$ & $\infty$ & $\sim 4.26$&$\sim 161$  \\ 
\hline $\pi/3$ & $\sim 10^6$ &$\sim0.31$ &$\sim 22400$\\ 
\hline $\pi/4$ & $\sim 10^5$ &$\sim1.81$ & $\sim 7570$\\ 
\hline $\pi/6$ & $\sim 22580$ & $\sim 2.81$ & $\sim 1510$\\ 
\hline $0$ & $\sim 7144 $ & $\sim 3.13$ & $\sim 103$\\
\hline
\end{tabular} 
\end{center}
\caption{Table 1: Entropy density of a Kerr-Newman black hole for different value of $\theta$, in $r_+$, $r_-$ and, for $r\rightarrow 0$ .}
\label{tabla}
\end{figure}

%\subsection{Kerr black holes's entropy density for new $P$ scalar}

In order to check if our definition of the $P$ scalar (Equation \ref{p2}) can be used in other non spherically symmetric space-times, we will test it in Kerr black holes.

The calculation of the $P$ scalar by Equation \ref{p2} yields:
\begin{equation}
P=M^2 \dfrac{48 \widetilde{p}_{1}\;\widetilde{p}_{2}\;\widetilde{p}_{3}}{(r^2+a^2\cos^2\theta)^6},
\end{equation}
where:
\begin{eqnarray*}
\widetilde{p}_{1} & = & r^2-4r\cos\theta a+a^2\cos^2\theta,\\
\widetilde{p}_{2} & = & r^2+4r\cos\theta a+a^2\cos\theta, \\
\widetilde{p}_{3} & = & a^2\cos^2\theta-r^2,\\
\end{eqnarray*}
and the entropy density gives:
\begin{equation}
\mathfrak{s}=\left|\dfrac{192r\left(10a^6\cos^6\theta-45a^4\cos^4\theta r^2+24r^4a^2\cos^2\theta-r^6\right)}{(r^2+a^2\cos^2\theta)^7}\right|.
\end{equation}

\begin{figure}[b]
 \begin{center}
  \includegraphics[height=5cm, width=8cm,]{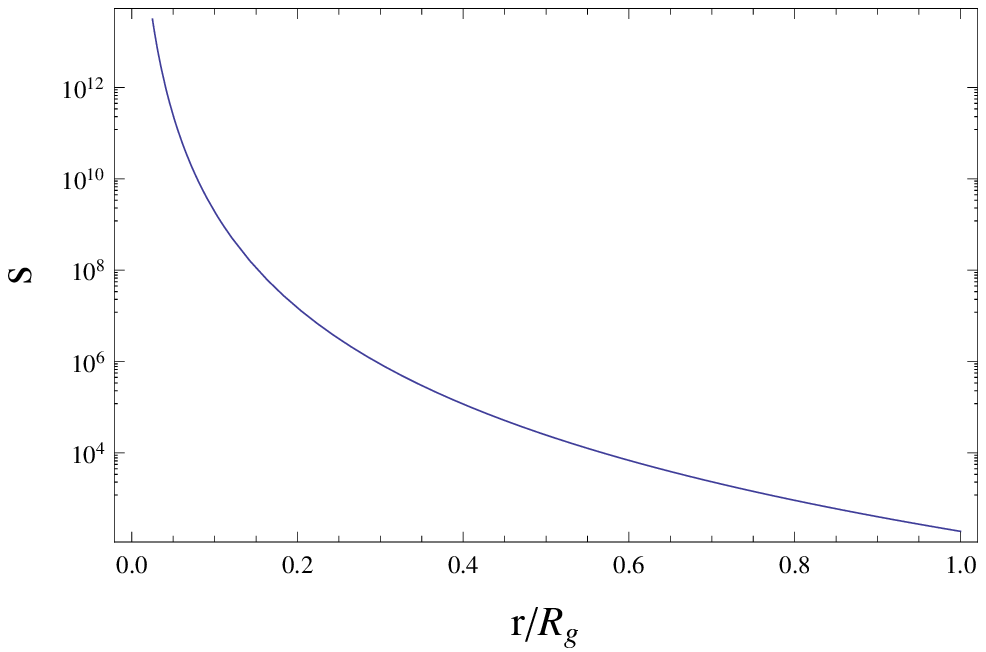}
 \end{center}
\caption{Entropy density of a Kerr black hole as a function of the radial coordinate for $P=C^{\alpha\beta\gamma\delta}C_{\alpha\beta\gamma\delta}$. Here, $c= G =k_{K} = 1$, $\theta=\pi/2$, and $a= 0.6$.}
\label{Kerrn1}
\end{figure}

\begin{figure}[b]
 \begin{center}
  \includegraphics[height=5cm, width=8cm,]{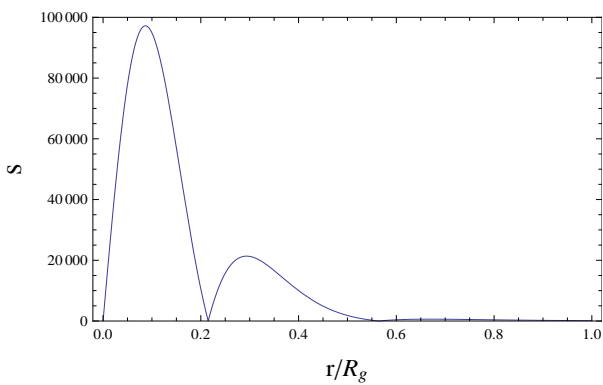}
 \end{center}
\caption{Entropy density of a Kerr black hole as a function of the radial coordinate for $P=C^{\alpha\beta\gamma\delta}C_{\alpha\beta\gamma\delta}$. Here, $c= G =k_{K} = 1$, $\theta=\pi/4$, and $a = 0.6$.}
\label{Kerrn2}
\end{figure}

Towards the center of the object the entropy density goes to zero, except for $\theta=\pi/2$ and $r=0$ where the ring singularity is located (see Figures \ref{Kerrn1} and \ref{Kerrn2}). We conclude that this new scalar also gives a reasonable description of the  gravitational entropy density in the Kerr space-time.

We remark that in axisymmetric space-times, the vector field $\vec \Psi$ could have an angular component:
\begin{equation}
\vec \Psi = P \left(e_{{\rm{r}}}+e_{{\rm{\theta}}}\right),
\end{equation}
where $e_{{\rm{\theta}}}$ is a unitary vector. According to Equation \ref{den}, the entropy density then takes the form:
\begin{equation}\label{den1}
\mathfrak{s}=k_{\rm{S}}\left|\frac{1}{\sqrt{-g}} \left[\frac{\partial}{\partial r}\left(\sqrt{-g} P\right)+\frac{\partial}{\partial \theta}\left(\sqrt{-g} P\right)\right]\right|.
\end{equation}
The second term of this equation is singular for $\theta = 0$ and $\theta = \pi$, that is, at the poles of both Kerr and Kerr-Newman black holes. The vector field $\vec \Psi$ is not regular over the whole horizon and consequently cannot be used as a reliable tool to estimate the gravitation entropy. This a direct consequence of the coupling of space and time components in rotating systems.
%\begin{equation}
%\mathfrak{s}= k_{\textit{\tiny{KN}}}  \nabla. (P\vec{e_{r}}).
%\end{equation}

%\begin{eqnarray}
%\mathfrak{s}&=& 2r\sqrt{\dfrac{A}{B}}\dfrac{\sqrt{\Delta}}{(\rho^{2})^{\frac{3}{2}}} +\sqrt {{\dfrac {A}{B}}}
%\dfrac{\left( \,r-\,M \right)}{\sqrt{\rho^{2}}} \left( \dfrac{1}{\sqrt {\Delta}}   - \dfrac{\sqrt {\Delta}}{\chi^{2}}  \right)\nonumber\\
%&+& \dfrac{1}{2}\, \dfrac{\sqrt{\Delta}}{\sqrt{\rho^{2}}}\left( {\dfrac {A'}{B}}-A{\dfrac { B' }{B ^{2}}} \right)\sqrt{\dfrac{B}{A}}, 
%\end{eqnarray}
\section{Conclusions}

The entropy density for the spherically symmetric compact objects, Reissner-Nordström black holes and wormholes, shows the expected features for a good measure of the gravitational entropy density in terms of Rudjord et. al.'s proposal: entropy density is everywhere well-defined, except in Reissner-Nordström black holes where the singularity is located, and goes to zero as the field decreases. In the case of a simple wormhole space-time, $\mathfrak{s} = 0$ at the throat because of the presence of exotic matter. For axisymmetric space-times, however, Rudjord et. al's formulation  has to be modified because the entropy density presents divergences. In this work we conclude that in the case of rotating bodies, the entropy density has to be calculated from the full metric form. Furthermore, the $P$ scalar is redefined as the Weyl scalar. These changes give a reasonable description of the gravitational entropy density: the function $\mathfrak{s}$ is well-defined everywhere except where space-time is singular and it tends asymptotically to zero outside the black hole. \\

We remark, however, that classical estimators as discussed here, should be tested in cosmological scenarios in order to determine their validity in more general space-times. A complete characterization of the gravitational entropy will only be possible with a quantum theory of gravitation, which should be a singularity-free theory. In the meanwhile, classical estimators can be helpful for some applications. The general validity of the Second Law of Thermodynamics in black hole interiors remains an open issue, that will be addressed in a forthcoming work.

\begin{acknowledgements}
Relativistic astrophysics with G. E. Romero is supported by CONICET through grant PIP 2010/0078 and the Spanish
Ministerio de Ciencia e Innovacion (MICINN) under grant AYA2010-21782-C03-01.
\end{acknowledgements}


\begin{thebibliography}{}

\bibitem{Boltz1}
Boltzmann, L.: Weitere Studien $\ddot{u}$ber das W$\ddot{a}$rmegleichgewicht unter Gasmolek$\ddot{u}$len. {\it Wiener Berichte} {\bf 66}, 275-370 (1872)

\bibitem{Boltz2}
Boltzmann, L.: $\ddot{U}$ber die Beziehung zwischen dem zweiten Hauptsatz der mechanischen W$\ddot{a}$rmetheorie und der Wahrscheinlichkeitsrechnung respektive den S$\ddot{a}$tzen $\ddot{u}$ber das W$\ddot{a}$rmegleichgewicht. {\it Wiener Berichte} {\bf 76}, 373-435 (1877)

\bibitem{Gold}
Gold, T.: The arrow of time. {\it Am. J. Phys.} {\bf 30}, 403-410 (1962)

\bibitem{p2}
Penrose, R.: {\it General Relativity, an Einstein Centenary Survey}. In: Hawking, S.W., Israel, W. (eds.) Singularity and Time-Asymmetry pp, 581-638. Cambridge Univ. Press (1979)

\bibitem{Gron}
Rudjord, $\O$, Gr$\o$n, $\O $., Sigbj$\o$rn, H.: The Weyl curvature conjeture and black hole entropy. {\it Phys. Scr.} {\bf 77} {\it Issue 5}, 055901, 1-7 (2008)

\bibitem{bek}
Bekenstein, J.D.: Generalized second law of thermodynamics in black holes. {\it Phys. Rev. D} {\bf 9}, 3292-3300  (1974)

\bibitem{Nord}
Nordström, G.: Uber die M$\ddot{o}$glichkeit, das elektromagnetische Feld und das Gravitationsfeld zu vereinigen. {\it Zeits. Phys.} {\bf 15}, 504-506 (1914)

\bibitem{Einstein}
Einstein, A.: Die Feldgleichungen der Gravitation. {\em Preussische Akademie der Wissenschaften}, 844-847 (1915)

\bibitem{p4} 
Penrose, R.: Gravitational Collapse: the role of general relativity. {\it Riv, Nuovo Cimento 1}, 252-276 (1969)


\bibitem{po-is}
Poisson, E., Israel, W.: Internal structure of black holes. {\it Phys. Rev. D} {\bf 41}, 1796-1809 (1990)


\bibitem{Gus}
G. E. Romero, {\it Lecture Notes on Introduction to Black Holes Astrophysics}, FCAG (UNLP) (2010)

\bibitem{Visser}
Visser, M.: {\it Lorentzian Wormholes}. AIP Press, New York (1996)

\bibitem{Morris}
Morris, M.S., Thorne, K.S.: Wormholes in spacetime and their use for intestellar travel: A tool for teaching general relativity. {\it Am. J. Phys.} {\bf 56}, 295-412 (1988)


\bibitem{Kerr}
Kerr, R.P.: Gravitational field of a spinning mass as an example of algebraically special metrics. {\it Phys. Rev. Lett.} {\bf 11}, 237-238 (1963)

\bibitem{Newman}
Newman, E. et. al.: Metric of a rotating, charged, mass. {\it J. Math. Phys.} {\bf 6}, 918-919 (1965)




%\bibitem{p1}
%Penrose, R.:{\it Proc. First Marcel Grossmann Meet. Gen. Rel. (ICTP
%Trieste)}. In Ruffini, R. (eds.) Space-time singularities. North-Holland, Amsterdam (1977)
%\bibitem{p3}
%Penrose, R.: On the second law of thermodynamics. {\it J. Stat. Phys.} {\bf 77}, 217-221 (1994)

\end{thebibliography}

{}

%\begin{thebibliography}{}
% etc
%\end{thebibliography}

\end{document}